# Ultrasensitive Detection of a Protein by Optical Trapping in a Photonic-Plasmonic Microcavity

*Miguel A. Santiago-Cordoba[1], Murat Cetinkaya[2], Svetlana V. Boriskina[3], Frank Vollmer[4*], Melik C. Demirel[1*]*

[1]Materials Research Institute, 212 EES Bldg, Pennsylvania State University, University Park, Pennsylvania 16802, USA.
Email: MDemirel@engr.psu.edu
[2]BASF SE, Carl-Bosch Strasse 38, Ludwigshafen, 67056, Germany.
[3]Mechanical Engineering Department, Massachusetts Institute of Technology, Cambridge, Massachusetts, 02139, USA
[4]Laboratory of Biophotonics and Biosensing, Max Planck Institute for the Science of Light, Guenther-Scharowsky-Str. 1, Erlangen, 91058, Germany. Email: frank.vollmer@mpl.mpg.de

Microcavity and whispering gallery mode (WGM) biosensors derive their sensitivity from monitoring frequency shifts induced by protein binding at sites of highly confined field intensities, where field strengths can be further amplified by excitation of plasmon resonances in nanoparticle layers. Here, we propose a mechanism based on optical trapping of a protein at the site of plasmonic field enhancements for achieving ultra sensitive detection in only microliter-scale sample volumes, and in real-time. We demonstrate femto-Molar sensitivity corresponding to a few 1000s of macromolecules. Simulations based on Mie theory agree well with the optical trapping concept at plasmonic 'hotspots' locations.

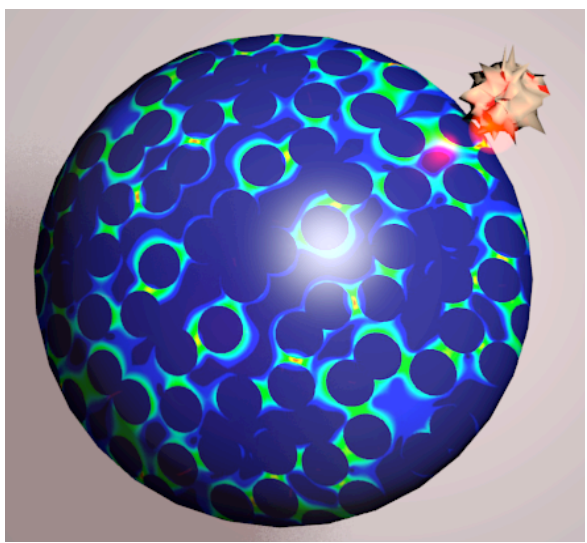

Protein detection with plasmon-coupled microcavity biosensor

# 1. Introduction

Label free optical biosensors enable the monitoring of biomolecules and their interactions in often highly sensitive diagnostic assays. For example, mechanical biosensors, such as atomic force microscopy[1,2] and quartz crystal microbalance[3], electrical biosensors such as impedance[4], and voltammetry[5,6], optical sensors such as Raman spectroscopy[7-11], plasmon resonance[12] and emerging Whispering Gallery Mode (WGM)[13-18] sensors have been utilized for detecting biomolecules at μg-ng/ml sensitivity levels. WGM biosensing offers a particularly sensitive approach to quantify the mass loading of biomolecules on the resonator surface with ultimate sensitivity estimated on the single molecule level[19,20]. WGM biosensors derive their unprecedented sensitivity from the use of high quality-factor (Q-factor) optical resonances to monitor wavelength shift signals upon binding of biomolecules or nanobeads to the resonator surface[20-25]. The simplest WGM biosensor is a glass microsphere (typically 50-100 μm in diameter) where the resonant light remains confined by total-internal reflection. The WGM biosensing approach has been shown to be highly sensitive down to the single virus or nanobeads level (~100 nm)[22,23,25,26] in purified buffer solutions. In complex samples such as blood plasma, recognition elements such as antibodies or DNA oligonucleotides immobilized at the WGM microsphere surface are needed for specific detection tasks[27,28].

Achieving the goal of single molecule detection of proteins (~10 nm in size) and their actions in solution requires mechanisms for boosting sensitivity in WGM biosensing[13]. Several approaches have been investigated recently, and specific examples for enhancement mechanisms include the use of multiplexed sensor arrays[27], self-referenced mode-splitting techniques[25,26] as well as hybrid photonic-plasmonic WGMs[29,30]. In the latter approach, a WGM is tuned close to the plasmon resonance of a surface bound gold (Au) nanoparticle (NP) 'antenna' so that a hotspot of high field intensity is created at the NP site – without significant loss of Q-factor. Molecules binding to the hotspot location can dramatically tune the WGM wavelength of such hybrid photonic-plasmonic detectors. In fact several orders of magnitude in sensitivity enhancements are predicted[29,31], bringing label free single molecule detection in aqueous solution and the study of single proteins 'in action' within reach. This hybrid photonic-plasmonic sensing concept was first demonstrated with an Au NP layer coupled to a WGM biosensor[29] where different amount of bovine serum albumin (BSA) proteins pre-adsorbed to the NP layer has been detected down to pM concentration levels. A drawback of this approach is the fact that measurements cannot be done directly in solution. Furthermore, real-time analysis is not possible since this method requires extraction, filtration and drying of the NP layer before probing with the WGM biosensor. In addition, proteins are adsorbed randomly within the NP layer (i.e., outside of plasmonic field enhancements sites), which lowers the sensitivity of the detection. Therefore, there is a need for a new approach for probing proteins in solution to achieve ultra-sensitive detection in WGM.

Here, we propose to overcome these problems by optical trapping of protein molecules at the sites of plasmonic field enhancements in a random Au NP layer. We implement real-time detection of protein adsorption from microliter-scale sample volumes and demonstrate ultra-sensitive detection down to fM solution concentration levels, corresponding to only a few thousand of protein molecules. Calculation of plasmonic field enhancements using electromagnetic Mie-theory combined with simulation of stochastic protein binding events suggest that indeed optical trapping of the proteins at highly sensitive plasmonic hotspot locations is essential for achieving high sensitivity in microcavity biosensing.

# 2. Experimental and Computational Methods

## 2.1 Computational Methods

*Generalized Mie Theory:* Both, far and near field spectra of hybrid plasmonic-photonic structures were calculated in the frame of the generalized multi-particle Mie theory, which provides an exact analytical solution of Maxwell's equations for an arbitrary configuration of $L$ spherical scatterers (NPs and/or microspheres)[32-34]. The total electromagnetic field scattered by the hybrid structure was constructed as a superposition of partial fields scattered by each scatterer:

$$\mathbf{E}_{sc}^{l} = \sum_{(n)} \sum_{(m)} \left( a_{mn}^{l} \mathbf{N}_{mn} + b_{mn}^{l} M_{mn} \right) \quad l = 1,...L \quad (1)$$

where $N_{mn}$ and $M_{mn}$ are spherical vector wave functions. A matrix equation for the Lorenz-Mie multipole scattering coefficients ($a^{l}_{mn}$, $b^{l}_{mn}$) was obtained by imposing the continuity conditions for the tangential components of the electric and magnetic fields on the surfaces of NPs and the microsphere, and by truncating the infinite series expansions to a maximum multipolar order $N$. The far field extinction spectra of a hybrid sensor were calculated under plane-wave illumination incident normally on the top on the microsphere. We used the experimentally measured Au refractive index values from Johnson and Christy[35] in the simulations. The Au NP clusters mor-phology (55nm NP diameters, 4 nm minimum interparticle gap) reproduced the experimental values determined by the scanning electron microscopy (Figure 1c).

*Monte Carlo (MC) Simulations:* Protein adsorption is a complex process and influenced by many surface properties such as surface wettability, chemical composition and morphology[36, 37]. Various adsorption models including Langmuir adsorption, virial expansion and scale particle theory were proposed to model proteins adsorption on surfaces[38]. Langmuir adsorption and virial expansion seldom predict the correct behavior due to conformation change of protein upon adsorption to the surface, and interaction with other adsorbed proteins on the surface. Scale particle theory incorporates excluded volume and shape effects, which predicts isotherm to be a function of protein shape. The isotherm broadens due to excluded volume, and steepens due to attractive interactions between proteins. However, the scaled particle theory is limited to well-defined geometries and does not incorporate multilayer adsorption. Computational models, such as molecular dynamics, and MC, have advantages compared to analytical model listed above, which solve major issues related to electrostatic and pH effects. However, molecular dynamics simulations are limited to single protein adsorption due to cost of the computational time. Cooperative adsorption of an ensemble of protein can be studied using MC simulation, which provides large number of configurations for adsorbed proteins with low computational cost[36]. It should be noted that stochastic approach of the MC simulation could be a disadvantage compared to the deterministic molecular dynamic simulation, which provides dynamical information.

The MC simulations are performed with a custom written code. The model is a two-dimensional square lattice simulation with periodic boundaries, where each lattice on the substrate is as large as a BSA protein (~4 nm radius). MC iterations (i.e. adsorption, desorption, and diffusion on the substrate) were performed according to the Metropolis sampling algorithm[39]. The probability of accepting an MC move is therefore:

$$P = \min\left[1, e^{-(U_{new}-U_{current})/\beta}\right] \quad (2)$$

where $\beta = 1/k_B T$ with $k_B$ being Boltzmann's constant and $T$ the absolute temperature, $U_{new}$ is the new potential energy of the system, and $U_{current}$ is the current potential energy of the system. Reduced units are used for simulation parameters. Under non-biased conditions (i.e. when the field effects are not involved), the adsorption energy of a single protein molecule was taken as -1.0 $k_B T$. Field effects are taken into account by biasing the probability of finding a non-occupied lattice and also by modifying the Metropolis acceptance criterion:

$$P_{bias} = \min\left[1, (f_{new}/f_{current}) \cdot e^{-(U_{new}-U_{current})/\beta}\right] \quad (3)$$

where $f_{new}$ and $f_{current}$ are the field intensity values at the corresponding locations of the substrate. These intensity values are the results of generalized Mie theory field patterns (see Figure 4a). The cooperative adsorption of the proteins is taken into account by adding an extra energy of -0.1 kT for each occupied next neighbor lattice. This value has been assigned after checking that the protein adsorption is moderately affected (i.e. Langmuir type of behavior is no longer observed), but the field effects are not suppressed either. Simulations are run for 50 million MC steps, after which the results are found to converge at low protein concentrations and under field effects. Each simulation is replicated 3 times with a different seed using the random number generator of the computer.

## 2.2 Experimental Methods

*Materials*: Chloroauric acid (HAuCl4), sodium citrate, (11-Mercaptoundecyl)-N,N,N-trimethylammonium bromide, Bovine Serum Albumin (BSA), and phosphate buffer saline (NaCl 0.138 M; KCl 0.0027 M; pH 7.2, at 25 °C after dissolving it in 1L ultrapure water) were purchased form Sigma-Aldrich. Anodic alumina oxide membranes (100 nm) were purchased from Whatman. Glass microfiber was purchased from Corning.

*Gold NP preparation: 1. Citrate method:* The synthesis of the Au-citrate NPs was done following a modification of the Lee and Meisel's method[40] as the one described by Kruszewski and coworkers[41]. Briefly, Au colloids were prepared by a reduction of $HAuCl_4$ ($10^{-3}$ M, 100 mL) with sodium citrate (1%, 4 mL). The $10^{-3}$ M $HAuCl_4$ solution was heated to 100 °C under constant stirring. Then, 4 mL of a 1% trisodium citrate solution was added to the reacting mixture. Two minutes after the addition of trisodium citrate the reacting mixture turned to a black color. The reaction was let boiling for 1 h, finally exhibiting a deep red color.

*Gold NP preparation: 2. Amino method:* The preparation of the Au-amino NPs was done in a similar way to the one described by Kunitake et al.[42] Basically, Au-citrate NPs were functionalized upon preparation with (11-Mercaptoundecyl)-N,N,N-trimethylammonium bromide (10 mg). After addition of the trimethylammonium thiol the solution changed to a brown-yellowish color, and the reaction was stopped when the solution turned to a dark brown-reddish color.

*Optical measurements:* WGM spectra were obtained with a tunable external cavity laser operating at ~633 nm nominal wavelength (TLB 7000, New Focus). A silica microsphere (radius: ~60-250 μm) coupled to a tapered optical fiber[29] (SMF-28, Thorlabs) was employed as the microcavity probe. The microspheres were fabricated by melting the tip of a piece of single mode optical fiber (SMF-28, Thorlabs) with a $CO_2$ laser [23]. Then the microsphere cavity was permanently aligned to the tapered optical fiber with the help of mechanical stages such that a WGM is excited along the equator of the microsphere, see Figure 1. The fiber-coupled microsphere was mounted on mechanical stages in such a way that it can probe via its evanescent field the NP layer when brought in contact. The WGM evanescent field has a considerable spatial extend with an approximate evanescent field length ~50nm in air and ~90nm in water, and excites plasmon resonances at several NP sites in the layer creating a random array of highly sensitive intensity hotspots. The plasmon-coupled WGM sensor is particularly sensitive to proteins binding at these hotspot locations where binding induces a large tuning of the plasmon-coupled WGM resonance wavelength, proportional to the field strength, E, encountered at the binding site, $r_v$[29],

$$\left(\frac{\Delta\lambda_r}{\lambda_r}\right) \cong \frac{\alpha/\varepsilon_0 |\mathbf{E}(\mathbf{r}_v)|^2}{2\int_V \varepsilon_r(\mathbf{r})|\mathbf{E}(\mathbf{r})|^2 dV} \quad (4)$$

where α is the excess polarizability of the protein molecule and the denominator represents the energy density integrated over the whole mode volume. The data is obtained by recording a baseline resonance wavelength signal, and then protein solution is added to the layer. After adding BSA solutions, the resonance wavelength shift Δλ is recorded until the signal saturates after approximately 10-15 minutes, dependent on protein concentration. Figure 1(**D**) shows such a shift in wavelength units [nm] for 0.1 pM concentration of BSA on the Au NP layer with amine modification. The wavelength shift measurements were normalized by dividing by the nominal laser wavelength (633 nm) and by multiplying with the microsphere radius R (in nm). The saturation value of this sphere normalized fractional shift R×Δλ/λ [nm] was plotted versus concentration in Figure 2. A different microsphere was used for each adsorption measurement.

*Filtration of Gold NPs:* Anodic anlumina oxide (AAO) 100 nm diameter membranes were soaked for 5 h in an ~5% solution of polyethyleneinimide (PEI) and ultra pure $H_2O$ (18.2 Ohm-cm). Then they were washed with ultra pure water, by soaking them in six consecutive baths (10 min each), and then they were dried overnight in a desiccator under vacuum. Au colloids previously prepared were suction filtered through the PEI-modified nanoporous membrane. Then, the Au NP layer template was vacuum dried for at least 4 h.

*Protein Measurements:* BSA adsorption experiments were performed at different concentrations, ranging from 0.001 pM to 1000 pM. First a wetted Au NP layer was brought into contact with the resonant microcavity. Then, 1.5 μL of BSA solution (phosphate buffered saline) was added to the Au NP membrane. After addition of the BSA, the solution wicks through capillary action into the membrane and delivers the BSA molecules to the sensor area where a wavelength shift is recorded as a function of time and analyzed using a custom Labview program that tracks the wavelength shift signal, see Figure 1(**D**).

## 3. Results and discussion

Figures 1A and 1B show the schematic for real-time label-free sensing platform of a BSA protein. The stable integration of the microsphere WGM biosensor with a wetted Au NP layer is critical for achieving ultra-sensitive detection. Therefore, the silica microsphere cavity remained fixed on the Au NP layer. The NP layer was kept wetted using a humidity chamber throughout the experiment. Initial Q-values of the microspheres are in the $10^6$-range. After stable coupling to the wetted NP layer the Q-factor drops slightly but remains in the $10^5$ range. Figure 1C shows the electron microscope image of Au NPs (approximately 55 nm in diameter) filtered on an aluminum oxide membrane forming a random distribution of NPs. Figure 1D shows a typical spectrum for the WGM biosensor. A stable resonance peak at specific wavelength λ is recorded before adding protein solution. Then, BSA protein, dissolved in PBS buffer, is added at microliter of sample volumes. The solution is introduced to the NP layer by capillary suction, and immediately probed by the WGM of the microsphere cavity. The resonance wavelength shift is recorded as a function of time (see also experimental methods). Figure 1E shows the experimental setup.

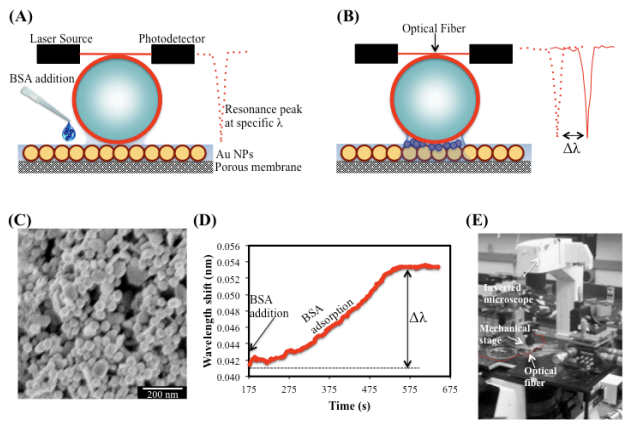

**Figure 1.** (**A**) Schematic diagram of the Au NP layer coupled to a WGM resonant microcavity. (**B**) Protein detection scheme by monitoring the wavelength shift of a resonant microsphere cavity. (**C**) Scanning electron micrograph of Au NPs immobilized onto an AAO nanoporous membrane. (**D**) Example of a BSA adsorption curve, upon BSA addition, the resonance spectra shift drastically and BSA adsorption is observed in real time as an increase in resonance wavelength. (**E**) A photograph of the WGM setup mounted on a high-precision stage located on an inverted microscope. The microscope helps to position and align the silica microspheres with respect to the fiber and the Au NP substrate.

Figure 2 shows microsphere normalized fractional wavelength shift of BSA adsorption as a function of BSA concentration ranging from fM to pM. We observe unexpectedly large resonance wavelength shifts with high sensitivity for WGM detection of BSA protein down to few 1000s of molecules. We design two experiments with NPs that have been modified with different surface monolayers. A negatively charged citrate and a positively charged amino group monolayer were chosen to study the effects of surface charge on nonspecific BSA protein adsorption from solution to the Au NP layer. We observed a slightly higher protein adsorption to Au NPs with amino groups. This is not surprising since BSA is negatively charged in solution at physiological pH, and it is known that BSA forms monolayers on monolayers of amino-silanized surfaces[43]. Table 1 shows the adsorption kinetics as a function of concentration (i.e. rate shift $R\Delta\lambda_{eq}/(\lambda \times t_{eq})$, where $\Delta\lambda_{eq}$ is the saturation shift in equilibrium and $t_{eq}$ is the time).

The level of sensitivity on the order of few 1000s of molecules (fM concentration levels) for both NP coatings is very surprising, and cannot be explained from random binding of the BSA molecules to the NP surface. Instead, we hypothesize that the protein molecules prefer to bind to hotspot locations (i.e. closely spaced random NPs) of plasmon resonances excited in the NP layer due to optical trapping. To validate our hypothesis, we calculated the electromagnetic field distribution in a model NP layer using generalized Mie theory (see computational methods) and simulated the expected wavelength shift due to the binding of proteins.

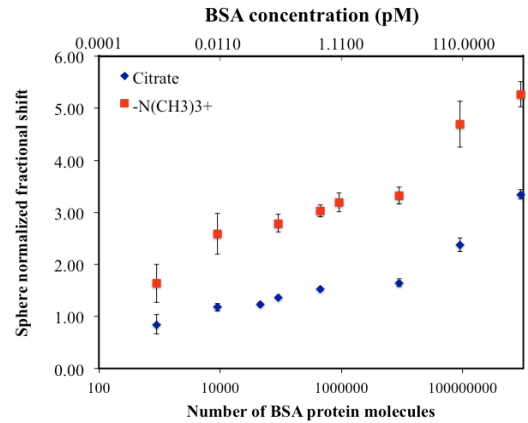

**Figure 2.** Normalized WGM shift measured at ~633 nm probing wavelength of the Au NP layer, see experimental methods section. The shift becomes larger as the number of BSA protein molecules added to the Au NP layer increases.

**Table 1**. Rate of the wavelength shift signal corresponding to the different adsorption kinetics of BSA measured in real time for citrate and amino modified Au NPs.

| Concentration (pM) of the droplet | Number of BSA molecules | Rate shift $R\Delta\lambda_{eq}/(\lambda \times t_{eq})$ in units of [nm/s] | |
|---|---|---|---|
| | | Citrate | Amino |
| 0.001 | ~$10^3$ | 0.00235 | 0.00487 |
| 0.01 | ~$10^4$ | 0.00244 | 0.00577 |
| 0.1 | ~$10^5$ | 0.00368 | 0.00930 |
| 0.5 | ~$5 \times 10^5$ | 0.00464 | 0.01017 |
| 10 | ~$10^6$ | 0.00474 | 0.01108 |
| 100 | ~$10^7$ | 0.00610 | 0.01340 |
| 1000 | ~$10^8$ | 0.00872 | 0.01756 |

We simulated model hybrid structures composed of a 5 μm diameter silica microsphere and finite-size planar NP clusters of 3 NPs to simplify the numerical analysis. Our previous work confirmed that such scaled model structures provide qualitative physical picture of the interactions between the evanescent fields of WGMs and plasmonic modes of NP arrays[29, 44]. Figure 3 shows the result of the calculations for a linear NP trimer with 4nm-wide inter-NP gaps coupled to the 5 μm silica microsphere through a 1nm-wide gap. Under external light illumination as schematically shown in Figures 3A and 3B, high-intensity electromagnetic hotspots are generated in the gaps of the NP cluster (Figure 3C) owing to the excitation and strong near field coupling of localized plasmon modes on Au NPs. The hot-spot intensities are boosted by orders of magnitude[29, 44-47] if the NP cluster is resonantly excited via the field of the WGM generated in the microcavity (compare Figures 3C and 3D)[1].

In Figure 3E, we plotted the fractional wavelength shift of the hybrid WG-plasmon mode of the sensor due to adsorption of BSA molecules to different spatial areas of the NP cluster. BSA molecules themselves are modeled as spherical NPs of 3.4nm radii and refractive index of 1.45. As shown in the inset, first four molecules were assumed to adsorb in the area of the highest field intensity between the Au NPs (blue dots), followed by the next four molecules (shown as red dots), which formed the second molecular layer in the cluster gaps. As can be seen in Figure 3D, the top-layer molecules are immersed in the lower-intensity region of the optical mode field. Clearly, molecules that adsorbed in the high-intensity hotspots cause larger wavelength shifts than those forming the second molecular layer, and so forth. Binding at high field intensities can therefore explain the large wavelength shifts that were measure in our experiments (see also experimental methods section). The particularly large wavelength shift that is expected for binding of the first molecules at highest field intensities might explain the offset seen in our measurements in Figure 2 indicating that we do not yet have a large enough signal-to-noise ratio to resolve these first binding events.

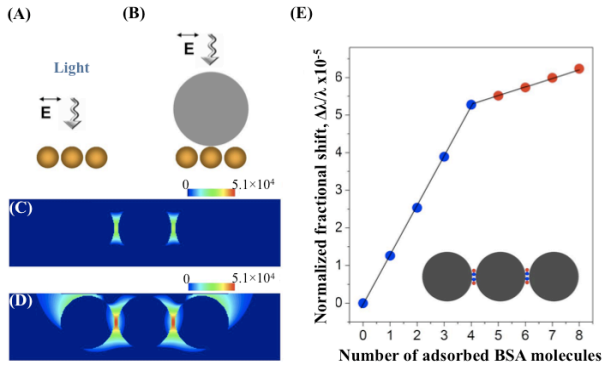

**Figure 3.** (**A**,**B**) Schematics of the Au NP cluster (trimer) (**A**) and a hybrid microsphere-NP cluster sensor (**B**) under the plane wave illumination. Au NP diameters are 55nm and interparticle separations are 4nm. Silica microsphere has a diameter of 5micron, and microsphere-NP separation is 1nm. (**C**,**D**) Calculated spatial near-field intensity distributions in the isolated (**C**) and microsphere-coupled NP cluster (**D**) at 634.9nm wavelength. (**E**) Fractional wavelength shifts of the hybridized WGM-plasmonic NP mode of the sensor caused by the adsorption of BSA molecules one-by-one. The inset shows the positions of the adsorbed molecules relative to the hot spots formed in the NP cluster (compare with (D)).

We also examined whether the calculated field strengths between closely spaced NPs can indeed promote the binding of BSA proteins by optical trapping. In general, the total time-averaged optical force acting on a molecule or a particle illuminated by incoming light can be obtained from the Maxwell's stress tensor[48]. However, for the molecules modeled as particles with the radii much smaller than the wavelength of incident light the total time-averaged optical force acting on the molecular particle can be calculated within the Rayleigh (dipole) approximation as the sum of the gradient force and the dissipative force[48-50]:

$$\langle \mathbf{F} \rangle = -\nabla \langle U \rangle + \langle F_D \rangle = \frac{I_0 n}{c\varepsilon_0}(\alpha' \nabla M + \alpha'' \mathbf{k} M) \quad (5)$$

Here, $\langle U \rangle$ is the optical potential, $n$ is the refractive index of the ambient medium, $I_0$ is the incident field intensity, $M$ is the local electric field intensity enhancement at the molecular particle position, and $\alpha = \alpha' + i\alpha''$ is the isotropic complex NP polarizability, which for the molecular particle of radius $R$ and permittivity $\varepsilon_p$ embedded in the medium with permittivity $\varepsilon_s$ is calculated as follows:

$$\alpha = 4\pi R^3 \varepsilon_0 \varepsilon_s \frac{(\varepsilon_p - \varepsilon_s)}{(\varepsilon_p + 2\varepsilon_s)} \quad (6)$$

For a BSA molecule with radius of 3.4 nm and refractive index $n_p = 1.45$ embedded in water ($n_s = 1.33$) polarizability $\alpha = 4.546 \times 10^{-37}$ [$A^2 s^4 kg^{-1}$].

Following equation (5), the dissipative optical force component, which acts in the direction of the incident light propagation, is proportional to the imaginary part of the molecular particle polarizability. Therefore, the effect of this component is typically considered negligible for investigating lateral movement and trapping of small transparent dielectric particles. Under the action of the gradient force, the molecular particle drifts toward the region of higher electric field intensity, where the induced dipole has the lowest potential energy. The probability to find a sufficiently mobile NP or a molecule at a spatial position **r** is[50]

$$P(\mathbf{r}, U) \propto P_0(\mathbf{r})\exp\{-\langle U(\mathbf{r})\rangle/k_B T\}, \quad (7)$$

where $k_B$ is the Boltzmann constant, $T$ is the temperature, and $P_0(r)$ is the corresponding probability without optical fields. A gradient force strong enough to overcome the Brownian motion needs to be generated in the trapping area. For the maximum incident field intensity of the WGM used in the experiment $I_0 = 2 \cdot 10^9 [W/m^2]$, we estimate the optical potential required for trapping a NP of 3.4 nm radius and refractive index $n=1.45$ in aqueous environment (normalized by $k_BT$ ($T= 300$ K)[48, 50] as $U[k_BT]=1.1026 \times 10^{-4}$M. Our generalized Mie theory calculations of the field distributions in the various random configurations of microsphere-coupled Au NP clusters with minimum particle separations of 4nm show that high-intensity hotspots generated in the clusters have high enough field intensity enhancement (M ~ 4-9x10$^4$) to provide optical potential strong enough to trap a BSA molecule.

---

[1] see supporting information

We have studied the effects of the NP diameter and the NP cluster size on the intensity of the localized hot spots generated in the cluster under resonant excitation by the WG mode field (see supplementary materials). The results of our calculations indicate that (analogously to the case of a single microcavity-coupled NP[29]) the hot spot intensity is governed by the spectral overlap between the localized SP mode in a cluster and a WG mode in the microsphere. Furthermore, coupling of various SP modes within clusters[49] reduces the hot spot intensity, which initially drops rapidly with the increase of the cluster size and then saturates in clusters over 10 NPs in size. Overall, the performed calculations demonstrate that hybrid microsphere-cluster structures with NP diameters in the range of 30-60nm provide the optical potential for trapping of small protein molecules at the chosen excitation wavelength (~ 633nm).

Having confirmed the ability of a WGM-coupled Au NP cluster to trap BSA molecules at the site of high plasmonic field intensities, we simulated the distribution of BSA protein binding throughout a random NP layer using MC simulations.

Figure 4 shows the monte-carlo (MC) simulations results of BSA protein adsorption on NPs with and without the optical field. Cooperative effect (i.e. the influence of pre-adsorbed proteins on the adsorption of proteins in solution to adjacent sites) among proteins during the adsorption period is also considered for the MC simulation. Figure 4A shows the field intensity map calculated using the Mie theory, where orange to red regions are hotspots. The field intensity data is chosen as an input for the MC simulation as described in the methods section. Figure 4B and 4C show the final configuration of the MC simulation for 50 and 500 proteins respectively. These results show that even for large number of binding events (i.e. in the given scenario of competing binding sites as well as slight cooperativity) BSA molecules accumulate first at high intensity hotspots supporting our hypothesis of large wavelength shifts due to optical trapping. Figure 4D shows the coverage of hotspots as a function concentration of proteins. Similarly, we simulated protein coverage as a function of NP diameter (i.e. 40, 50 and 55nm) with the optical trapping (see supporting information). Switching off the field leads to random distribution of BSA molecules on the NP surface, where the coverage of hotspots is significantly lowered compared to optical field is on (i.e. optical trapping).

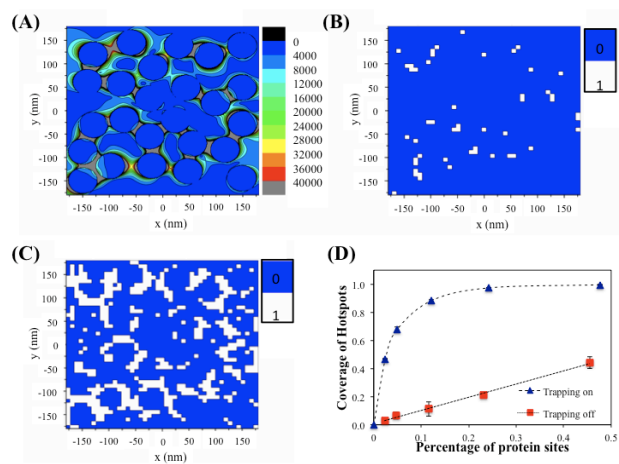

**Figure 4.** (**A**) The field intensity map for random Au NP is shown. This map is calculated using the generalized Mie-theory. Representative MC simulations results for N=50 and N=500 are shown in (**B**) and (**C**) respectively. BSA coverage, at the hotspot locations, shows clearly a different trend for light field on (blue) and light field off (red) MC simulations (**D**).

## 4. Conclusion

We have demonstrated a first hybrid photonic-plasmonic WGM biosensor that measures protein binding down to fM solution concentration levels. The unprecedented sensitivity towards detection of proteins is explained by optical trapping of proteins at highly sensitive plasmonic hotspots on a gold NP layer that is coupled to the WGM biosensor. This approach indicates a promising route towards achieving single molecule resolution in WGM biosensors coupled to engineered or random plasmonic nanoantennas. The approach of using a random NP layer has the advantage of integration to a microfluidic device with the added advantage of using gold NPs, which can be easily functionalized with recognition elements such as oligonucleotides or proteins. Sample analysis based on the hybrid photonic-plasmonic approach is rapid since detection is potentially sped up through the use of optical forces that deliver proteins to the sites of high field intensity. With the challenges of surface functionalization, detection speed and microfluidic integration addressed, we believe that highly sensitive hybrid photonic-plasmonic WGM biosensors in NP layers will have broad applications across multiple disciplines and areas including medical biosensing and drug screening[51, 52].

### Acknowledgements
We gratefully acknowledge financial support for this work from Burroughs Wellcome Fund (MSC), the Pennsylvania State University (MCD) and Max Planck Society (FV). We thank Prof. David Allara for useful discussions and comments.

### Author Contributions
MCD and FV planned and supervised the research. MSC carried out the WGM experiments. SB and MC developed the computational model for electromagnetic and MC calculations respectively. All authors contributed to writing and revising the manuscript, and agreed on its final contents.

## AUTHOR BIOS

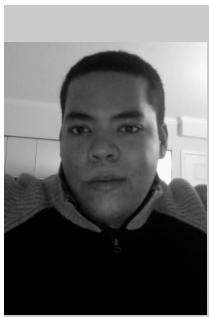

Miguel A. Santiago Cordoba is a Ph.D. candidate under the guidance of Prof. Melik Demirel at The Pennsylvania State University. His research is focused on understanding physicochemical properties of surfaces, and synthesis of novel materials for biosensor development. He received his B. Sc. Degree from the University of Puerto Rico – Rio Piedras in 2009.

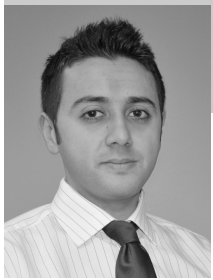

Dr. Murat Cetinkaya is a research scientist in the Materials Modeling Department at BASF SE, Germany. Dr. Cetinkaya's research area covers the molecular and continuum modeling of polymeric systems. He received Ph.D.(2008) and MSc (2005) degrees from The Pennsylvania State University under the guidance of Prof. Melik Demirel. He has been a postdoctoral researcher at Heidelberg University (2009) and The Heidelberg Institute for Theoretical Studies (2010) in Germany. He is currently working on multi-scale modeling of polymer and surfactant based formulations in product development.

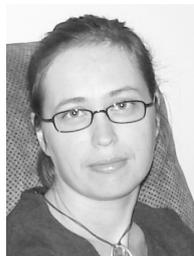

Dr. Svetlana V. Boriskina is a research associate at the Mechanical Engineering Department at Massachusetts Institute of Technology. Her research interests include nanophotonics, plasmonics and metamaterials for biosensing, information processing and renewable energy generation. She received her M.Sc. (Summa Cum Laude) and Ph.D. degrees in Physics and Mathematics from Kharkov National University (Ukraine). Dr. Boriskina is a holder of the Joint Award of the International Commission for Optics and the Abdus Salam International Centre for Theoretical Physics, a Senior Member of IEEE, a Member of OSA, and a Member of MRS.

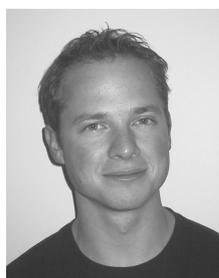

Dr. Frank Vollmer is leader of a Max Planck Research Group at the Max Planck Institute for the Science of Light in Erlangen, Germany. His Laboratory of Biophotonics & Biosensing explores the physics of optical biosensors and their application for clinical diagnostics on chip-scale devices. Dr. Frank Vollmer obtained his PhD from the 'Center for Studies in Physics & Biology', Rockefeller University, NYC, USA, in 2004. He then became leader of an independent research group at the Rowland Institute at Harvard University where he was appointed Rowland Fellow from 2004 to 2009. From 2010-2011, before joining the Max Planck Society, he joined the Wyss Institute for Bio-Inspired Engineering at Harvard University as Scholar-in-Residence

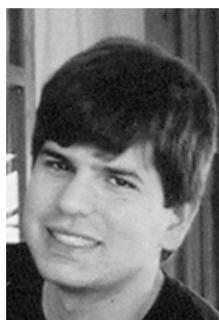

Prof. Melik Demirel is a tenured associate professor in the Engineering at The Pennsylvania State University. Prof. Demirel's research focuses on understanding the physicochemical properties of biological and synthetic materials. He received a Ph.D. in Materials Science and Engineering from Carnegie Mellon University (2002), and M.Sc (1998) and B.Sc (1996) degrees in Engineering from Bogazici University in Turkey. Prof. Demirel's achievements have been recognized, in part, through his receipt of a Young Investigator Award, an Alexander von Humboldt Fellowship, an Institute for Complex Adaptive Matter Junior Fellowship, the Pearce Development Professorship at Penn State.